

\font\rmu=cmr10 scaled\magstephalf
\font\bfu=cmbx10 scaled\magstephalf

\font\it=cmti10 scaled \magstephalf

\rmu

\font\rmus=cmr8
\font\rmuss=cmr6
\font\mait=cmmi10 scaled\magstephalf
\font\maits=cmmi7 scaled\magstephalf
\font\maitss=cmmi7
\font\msyb=cmsy10 scaled\magstephalf
\font\msybs=cmsy8 scaled\magstephalf
\font\msybss=cmsy7
\font\bfus=cmbx7 scaled\magstephalf
\font\bfuss=cmbx7
\font\cmeq=cmex10 scaled\magstephalf

\textfont0=\rmu
\scriptfont0=\rmus
\scriptscriptfont0=\rmuss

\textfont1=\mait
\scriptfont1=\maits
\scriptscriptfont1=\maitss

\textfont2=\msyb
\scriptfont2=\msybs
\scriptscriptfont2=\msybss

\textfont3=\cmeq
\scriptfont3=\cmeq
\scriptscriptfont3=\cmeq

\newfam\bmufam  \textfont\bmufam=\bfu
      \scriptfont\bmufam=\bfus \scriptscriptfont\bmufam=\bfuss

\hsize=15.5cm
\vsize=21cm
\baselineskip=16pt   
\parskip=12pt plus  2pt minus 2pt

\def\a{\alpha}
\def\b{\beta}

\def\g{\gamma}

\def\semi{\bigcirc\kern-1em{s}\;}

\def\del{\partial}
\def\ni{\noindent}
\def\R{{\rm I\!R}}

\def\one{{\mathchoice {\rm 1\mskip-4mu l} {\rm 1\mskip-4mu l}
{\rm 1\mskip-4.5mu l} {\rm 1\mskip-5mu l}}}
\def\Q{{\mathchoice
{\setbox0=\hbox{$\displaystyle\rm Q$}\hbox{\raise 0.15\ht0\hbox to0pt
{\kern0.4\wd0\vrule height0.8\ht0\hss}\box0}}
{\setbox0=\hbox{$\textstyle\rm Q$}\hbox{\raise 0.15\ht0\hbox to0pt
{\kern0.4\wd0\vrule height0.8\ht0\hss}\box0}}
{\setbox0=\hbox{$\scriptstyle\rm Q$}\hbox{\raise 0.15\ht0\hbox to0pt
{\kern0.4\wd0\vrule height0.7\ht0\hss}\box0}}
{\setbox0=\hbox{$\scriptscriptstyle\rm Q$}\hbox{\raise 0.15\ht0\hbox to0pt
{\kern0.4\wd0\vrule height0.7\ht0\hss}\box0}}}}
\def\C{{\mathchoice
{\setbox0=\hbox{$\displaystyle\rm C$}\hbox{\hbox to0pt
{\kern0.4\wd0\vrule height0.9\ht0\hss}\box0}}
{\setbox0=\hbox{$\textstyle\rm C$}\hbox{\hbox to0pt
{\kern0.4\wd0\vrule height0.9\ht0\hss}\box0}}
{\setbox0=\hbox{$\scriptstyle\rm C$}\hbox{\hbox to0pt
{\kern0.4\wd0\vrule height0.9\ht0\hss}\box0}}
{\setbox0=\hbox{$\scriptscriptstyle\rm C$}\hbox{\hbox to0pt
{\kern0.4\wd0\vrule height0.9\ht0\hss}\box0}}}}

\font\fivesans=cmss10 at 4.61pt
\font\sevensans=cmss10 at 6.81pt
\font\tensans=cmss10
\newfam\sansfam
\textfont\sansfam=\tensans\scriptfont\sansfam=\sevensans\scriptscriptfont
\sansfam=\fivesans
\def\sans{\fam\sansfam\tensans}
\def\Z{{\mathchoice
{\hbox{$\sans\textstyle Z\kern-0.4em Z$}}
{\hbox{$\sans\textstyle Z\kern-0.4em Z$}}
{\hbox{$\sans\scriptstyle Z\kern-0.3em Z$}}
{\hbox{$\sans\scriptscriptstyle Z\kern-0.2em Z$}}}}

\newcount\foot
\foot=1
\def\note#1{\footnote{${}^{\number\foot}$}{\ftn #1}\advance\foot by 1}

\def\frac#1#2{{#1\over #2}}
\def\text#1{\quad{\hbox{#1}}\quad}

\font\ch=cmbx12 scaled\magstephalf
\font\ftn=cmr8 scaled\magstephalf

\font\it=cmti10 scaled\magstephalf

\font\titch=cmbx12 scaled\magstep2
\font\titname=cmr10 scaled\magstep2
\font\titit=cmti10 scaled\magstep1
\font\titbf=cmbx10 scaled\magstep2

\nopagenumbers


\line{\hfil DFF 223/03/95}
\line{\hfil Mar 23, 1995}
\vskip4cm
\centerline{\titch QUANTUM ASPECTS OF 2+1 GRAVITY}
\vskip2.3cm
\centerline{\titname R. Loll\note{Supported by the European Human
Capital and Mobility program on ``Constrained Dynamical Systems"}}
\vskip.5cm
\centerline{\titit Sezione INFN di Firenze}
\vskip.2cm
\centerline{\titit Largo E. Fermi 2}
\vskip.2cm
\centerline{\titit I-50125 Firenze, Italy}

\vskip3.8cm
\centerline{\titbf Abstract}
We review and systematize recent attempts to canonically quantize
general relativity in 2+1 dimensions, defined on space-times
$\R\times\Sigma^g$, where $\Sigma^g$ is a compact Riemann surface
of genus $g$. The emphasis is on quantizations of the
classical connection formulation, which use Wilson loops as their
basic observables, but also results from the ADM formulation are
summarized. We evaluate the progress and discuss the possible quantum
(in)equivalence of the various approaches.

\vfill\eject
\footline={\hss\tenrm\folio\hss}
\pageno=1


\line{\ch 1 Introduction\hfil}

The aim of this article is to review and systematize attempts to
canonically quantize general relativity in 2+1 space-time dimensions.
We limit the scope to the case of pure Einstein gravity, possibly with
a non-vanishing cosmological constant $\Lambda$, on three-dimensional
manifolds of the form $\R\times\Sigma^g$, where $\Sigma^g$ is a compact
Riemann surface of genus $g\geq 1$. The inclusion of point sources
or matter, the possibility of topology changes, and lattice approaches
will not be
discussed. Basic knowledge of the classical theory and its geometric
interpretation will be presupposed, and wherever necessary we will
refer to the appropriate references for details. Our emphasis will
be on a comparison between the different existing quantization
methods, and their possible quantum (in)equivalence. This includes
issues like the degree of the classical reduction, the choice of the
basic variables to be quantized, and the role of the mapping class
group (the ``large diffeomorphisms").

If one wants to regard 2+1 gravity as a model for the
(3+1)-dimensional theory, with the local physical excitations
substituted by a finite number of topological degrees of freedom, it
is a somewhat embarrassing fact that many of the quantizations
proposed  have hardly progressed beyond the genus-one case, where
the spatial manifold $\Sigma$ is a torus $T^2$. On the one hand, one
finds a rich structure already in this case, but on the other one
knows that its mathematical structure is not very representative of
(and rather simpler than) the general higher-genus case. This should
be kept in mind when drawing conclusions about canonical quantum
gravity in general.

We shall distinguish between two different (but classically
essentially equivalent) approaches to 2+1 gravity, i) the geometric
description in terms of the Lorentzian three-metric ${}^{(3)}g_{\mu
\nu}$, and ii) a gauge-theoretic description in terms of a connection
one-form ${}^{(3)} A_\mu^a$,
taking values in an appropriate gauge algebra.
The first one is the three-dimensional analogue of the well-known ADM
framework, and involves the choice of a parameter representing
physical time. In the gauge-theoretic approach, no such choice is
required, and one has the option between a Chern-Simons formulation
(with a non-compact gauge group) and a closely related
three-dimensional version of the Ashtekar formulation of general
relativity. In all of these descriptions, 2+1 gravity takes the form of
a canonical system with first-class constraints \`a la Dirac. What
distinguishes it from other constrained field theories is the fact
that one can solve (part of) the constraints already classically to
the extent that the remaining reduced phase space is
finite-dimensional. This reduces the quantization problem to a
quantum mechanical one. Still, as we will see, there is no
unique way of setting up a quantum theory. Different proposals have
been made, depending on the classical starting point, and on various
requirements one may wish to impose on the quantum theory.
Another  special feature (and a consequence of the invariance under
space-time diffeomorphisms) is the absence of an a-priori defined
time parameter, leading to further ambiguities in the quantization.

In the familiar geometric approach, Einstein gravity with a
cosmological constant is defined by the Lagrangian

$$
S[g]=
\int d^3 x\sqrt{ -{}^{(3)}g } ({}^{(3)} R-2\Lambda)=\int dt\int_\Sigma
d^2 x(\pi^{ij}\dot g_{ij}-N^i{\cal H}_i-N {\cal H}),
\eqno(1.1)
$$

\ni where a (2+1)-decomposition \`a la ADM has been
performed to arrive at the second expression,
and the canonically conjugate variable pairs consist of the
spatial metric $g_{ij}(x)$ and the momentum $\pi^{ij}(x)$ on the
surfaces $t=const$, that depends on the extrinsic curvature $K^{ij}$
via $\pi^{ij}=\sqrt{{}^{(2)}g} (K^{ij}-g^{ij} K)$. (In this paper, we
will use $\mu$, $\nu,\ \dots$ for space-time indices, $i$, $j,\
\dots$ on two-dimensional spatial slices, and $a$, $b,\ \dots$
for internal indices.) The spatial
diffeomorphism and Hamiltonian constraints are given by

$$
\eqalign{
&{\cal H}_i=-2\nabla_j\pi^j{}_i=0\cr
&{\cal H}=\frac{1}{\sqrt{{}^{(2)} g}} g_{ij}g_{kl}(\pi^{ik}\pi^{jl}-
\pi^{ij}\pi^{kl})-\sqrt{{}^{(2)}g} (R-2\Lambda)=0,}\eqno(1.2)
$$

\ni (multiplied in the Lagrangian by the shift and lapse functions
$N^i$ and $N$), and thus have the same functional form as in the
(3+1)-dimensional case.

\vskip1.5cm
\line{\ch 2 Quantization in the connection formulation\hfil}

Since it was the gauge-theoretic reformulation of 2+1 gravity
and the demonstration of its ``exact solubility" by Witten [1] that
sparked off much of the recent interest in the theory, we will deal with
this case first. It is a remarkable fact that in both three and four
space-time dimensions, Einstein-Hilbert gravity may be reformulated
on a Yang-Mills phase space. How this leads to the Ashtekar
formulation of 2+1 and 3+1 and the Witten formulation of 2+1 gravity
has been discussed by Bengtsson [2].

Witten rewrites the action for 2+1 gravity as that of
a Chern-Simons theory for the Poincar\'e group $ISO(2,1)$ in three
dimensions, the group $SO(2,2)$ and $SO(3,1)$ for the cases
$\Lambda=0$, $\Lambda <0$ and $\Lambda >0$ respectively [1]. To this
end, one rearranges the dreibein $e_\mu^a$ and the
spin connection $\omega_\mu{}^a{}_b$ of the geometric first-order
formulation of the Einstein-Hilbert action
into a gauge algebra-valued connection form
$A_\mu=e_\mu^a P_a+\omega_\mu^a J_a$. (Recall that the three-metric in
this ansatz is the derived quantity $g_{\mu\nu}=e_\mu^a e_\nu^b
\eta_{ab}$, where $\eta$ is the three-dimensional Minkowski metric.)
The algebra generators $P_a$ and $J_a$ fulfil

$$
[J_a,J_b]=\epsilon_{abc} J^c,\quad
[J_a,P_b]=\epsilon_{abc} P^c,\quad
[P_a,P_b]=\Lambda\,\epsilon_{abc} J^c,\eqno(2.1)
$$

\ni where it is understood that internal indices are raised and lowered
using $\eta_{ab}$. The Lagrangian in this approach becomes

$$
\eqalign{
S'[{}^{(3)} A]=&\frac12 \int d^3 x\, \epsilon^{\mu\nu\lambda} {\rm Tr}
( A_\mu \del_\nu  A_\lambda +\frac13 \,  A_\mu [
A_\nu, A_\lambda ])\cr
=&\int dt\int_\Sigma d^2 x \,
\epsilon^{ij} (-e_{ia}\dot \omega_j^a +\lambda_A\,
F[{}^{(2)}A]^A_{ij}),}
\eqno(2.2)
$$

\ni where a 2+1 decomposition has been performed in the second step,
the index $A$ labels the six generators $T_A=(J_a,P_a)$ and $F$ is the
two-dimensional field strength of the spatial part of the connection
${}^{(3)} A$, taking values in the algebra of
the appropriate gauge group
$G$, where $G=ISO(2,1)$, $SO(3,1)$ and $SO(2,2)$, depending on the
value of the cosmological constant $\Lambda$. (Note that the
$\Lambda$-dependence is implicit in the commutators $[A,A]$.) For
$\Lambda=0$, this action is equivalent (at least for non-degenerate
dreibeins) to the three-dimensional Ashtekar action [3]

$$
S''[e,\omega]=\frac12 \int d^3 x\, \epsilon^{\mu\nu\lambda}  e_{\mu a}
F[{}^3 \omega]_{\nu\lambda}^a =\int dt\int_\Sigma d^2 x\,\epsilon^{ij}
(-e_{ia}\dot\omega_j^a +\mu_a {\cal D}_i e_j^a +\nu_a
F[{}^{(2)}\omega]_{ij}^a),\eqno(2.3)
$$

\ni where $F[{}^{(2)}\omega]$ is now the
two-dimensional field strength of the
$SO(2,1)$-connection, and $\cal D$ its covariant derivative. (This may
be considered as a special case of the equivalence between a Palatini
action for gauge group $G$ and a Chern-Simons action for the
corresponding inhomogeneous group $IG$ [4].) Defining
$E^{bj}=\epsilon^{jk} e_k^b$, we obtain in both cases a phase space
with symplectic structure

$$
\{\omega_{ia}(x),E^{jb}(y)\}=\delta_i^j\delta_a^b\delta^2 (x-y)
\eqno(2.4)
$$

\ni (that is, six variable pairs per space point). The Hamiltonian
densities are linear combinations of six first-class constraints each.
In the Chern-Simons formulation, they constrain the spatial component
of the field strength of the six-dimensional gauge algebra to vanish,
i.e. the corresponding connection $A$ to be flat, whereas in the
Palatini formulation they are made up of three Gauss law constraints
for $SO(2,1)$ and a flatness condition on the spin connection with its
{\it three}-dimensional gauge algebra. In the first case, the physical,
classically reduced phase space can therefore be identified with the
space of flat $G$-connections modulo $G$-gauge transformations. For
$\Lambda =0$, the alternative Palatini description yields a physical
phase space that is
a cotangent bundle over the reduced configuration space of
flat $SO(2,1)$-connections modulo $SO(2,1)$-gauge transformations. This
phase space coincides with the one of the Chern-Simons formulation for
$G=ISO(2,1)$, because the group manifold of $ISO(2,1)$ is isomorphic
to that of $T^*SO(2,1)$. Note that for non-vanishing $\Lambda$ the
reduced phase spaces are {\it not} cotangent bundles.

For the quantization of these reduced phase spaces it is important to
know concrete representations of the abstract quotient spaces, and
to decide which classical observables are to be carried over to the
quantum theory. Witten in his original paper envisaged a
quantization based (for $g\geq 2$) on a set of $2g$ $G$-valued
holonomy variables $U_i,\,V_i,\, i=1\dots g$, corresponding to the $2g$
generators $\a_i,\, \b_i,\, i=1\dots g$ of the homotopy group
$\pi_1(\Sigma)$,

$$
U_i=P\,\exp \oint_{\a_i} A^F,\qquad V_i=P\,\exp \oint_{\b_i} A^F,
\eqno(2.5)
$$

\ni where $A^F$ is a flat $Lie(G)$-valued connection one-form
on a spatial slice. The
$2g$ holonomies contain all information necessary for
reconstructing the moduli space of flat connections modulo gauge,
and can be interpreted as gluing data for simply connected patches
of Minkowski, de Sitter and anti-de Sitter space, corresponding to
$G=ISO(2,1)$, $SO(3,1)$ and $SO(2,2)$ (see, for example, [5]).
However, they are not free parameters, but i) are subject to
residual gauge transformations $U_i\rightarrow g U_i g^{-1}$,
$V_i\rightarrow g V_i g^{-1}$ at some arbitrary common base point of
the $\a_i$ and $\b_i$ in $\Sigma$, and ii) must obey a relation

$$
U_1V_1U_1^{-1}V_1^{-1}\dots U_gV_gU_g^{-1}V_g^{-1}=\one,\eqno(2.6)
$$

\ni which comes from the analogous defining relation among the
homotopy generators. In terms of these variables, the counting of
physical degrees of freedom is therefore $2g\times
dim(G)-dim(G)-dim(G)= (2g-2)\times dim(G)$.

Thus we may use this parametrization to describe the reduced phase
spaces of the Chern-Simons theory for the group $G$. In the case of
vanishing cosmological constant, we have the additional option of
describing the reduced {\it configuration} space by holonomy
variables, setting $G=SO(2,1)$ (because in this case there exists a
natural division between coordinates and momenta, a ``polarization"
of phase space).

Concrete quantization proposals have gone still one step further, and
based themselves on a set of explicitly gauge-invariant variables,
so-called Wilson loops, obtained from the holonomies by taking
traces,

$$
T(\g)[A^F]:={\rm Tr}\, P\,\exp\oint_\g A^F,\eqno(2.7)
$$

\ni where $\g$ is some element of $\pi_1 (\Sigma)$. The Wilson loops
are well known from their role as observables in usual gauge field
theories (see, for example, [6]). However, some
problems stand in the way of using them as ``coordinates"
on the physical moduli spaces. Note first that it is not enough to
use only the Wilson loops of the fundamental generators, since this
would give us just $2g$ degrees of freedom (out of the $(2g-2)\times
dim(G)$ needed). One thus has to include Wilson loops for more general
elements of $\pi_1 (\Sigma)$. However, these in general are subject
to certain algebraic constraints, the so-called Mandelstam
identities. In addition, there will be constraints on the Wilson
loops coming from the defining relation (2.6).

Another question is that
of the completeness of the Wilson loops, i.e. whether they separate
all points of the moduli spaces (so that no two physically
distinguishable $A$-configurations share the same values for all
Wilson loops). This is believed to be the case for the compact gauge
groups that typically appear in gauge theory, but not a priori clear
for the non-compact groups used in gravitational applications,
although it has been shown that for gauge group $SO(2,1)$ the traced
holonomies are essentially complete [3].  A related problem has been
investigated in the Ashtekar formulation of 3+1 gravity, where the
gauge group is $G=SL(2,\C)$ [7]. There also is the possibility of the
existence of inequalities between the Wilson loops, although this is
known not to happen for the hyperbolic sector of the
$G=SO(2,1)$-moduli space [8]. A problem of incompleteness seems to
occur for the case $\Lambda >0$, where the gauge group is $SO(3,1)$,
as has been remarked by various authors [9-12].

Another subtlety arises when one tries to make connection with the
geometric formulation in terms of a positive definite spatial
metric $g_{ij}$.
It turns out that the physical phase spaces there correspond to a
certain subsector of the moduli spaces
introduced above. (Other sectors also contain solutions to Einstein's
equations, but do not have the ``correct" signature for
$g_{ij}$, i.e. in general will allow for closed timelike curves.)
For example, for $G=SO(2,1)$, those are the configurations where all
holonomies are ``hyperbolic", i.e. ${\rm Tr}\,U_i >2$, ${\rm
Tr}\,V_i >2$ in the two-dimensional representation, which (for
$g\geq 2$) together form the so-called Teichm\"uller space ${\cal
T}(\Sigma^g)$ associated with the Riemann surface $\Sigma^g$. Since
they represent the true gravitational degrees of freedom, the most
obvious strategy is to quantize only them, ignoring the remaining
sectors of the moduli spaces. For $g=1$, which is qualitatively
different from the higher-genus case, the different sectors are
connected (they are not for $g>1$), and the structure of the
Chern-Simons moduli spaces is non-Hausdorff and rather complicated.
For $\Lambda=0$ this has been investigated in detail by Louko and
Marolf [13], and the analysis has been extended to $\Lambda\not= 0$
by Ezawa [14]. The former authors have also suggested a unified
quantum theory, in which all sectors are quantized together, and
which contains operators that map in between the different sectors.
-- From now on, when talking about moduli spaces, we will always mean
the appropriate geometrodynamic sector.

This in turn raises a question on prospective
``gauge-theoretic" quantum theories for 2+1 gravity: which feature of
the quantum representation reflects the fact that the correct, metric
sector is described? More generally, what indicates that the
underlying gauge group is $SO(3,1)$, $SO(2,2)$, rather than the
compact group $SO(4)$, say? The non-compactness of the gauge group is
responsible for a number of subtleties that occur in the existing
quantization programs. The two main lines of research are that of
Nelson, Regge and Zertuche on the one hand [16,24-28]
and Ashtekar et al, Smolin, Marolf and Loll on the other [3,19-21].
The former are mostly (but not only)
concerned with the case $\Lambda <0$, and a quantization based on a
finite subalgebra of Wilson loops, whereas the latter treat the case
$\Lambda=0$, and aim at a rigorous realization of the loop
quantization program, first proposed by Rovelli and Smolin for 3+1
gravity [15]. In both cases, explicit descriptions of the quantum
theory are available for $g=1$ and (incompletely) for $g=2$. We will
describe their main results in turn.
\vskip1cm
\line{\titit 2.1 Vanishing cosmological constant\hfil}

Let us begin with the case of a vanishing cosmological constant,
$\Lambda =0$. As explained above, it is special because the reduced
phase space is of the form of a cotangent bundle $T^*({\cal
A}^F/G)$ over the space of flat $SO(2,1)$-connections modulo gauge.
We will from now on work with the defining  two-dimensional
representation of $PSU(1,1)=SU(1,1)/\Z_2$, which has been used in
most applications. (We in any case are glossing over
differences that arise between using $G$ and some covering group
$\tilde G$.) In close analogy with the 3+1 theory, one introduces as
a convenient (over)complete set of phase space variables the
(normalized) Wilson loops [3]

$$
T^0(\g)[\omega]=\frac12\, {\rm Tr}\, U_\g,\qquad
T^1(\g)[\omega,E]=\oint_\g d\g^i\epsilon_{ij}{\rm Tr}\,(E^j U_\g),
\eqno(2.8)
$$

\ni with $\g\in\pi_1 (\Sigma^g)$, and where the canonical pairs
$(\omega,E)$ by slight abuse of notation denote now the coordinates of
$T^*{\cal A}^F$.

\ni The generalized Wilson loops $T^I$, $I=0,1$, form a closed Poisson
algebra with respect to the canonical structure induced on
$T^*{\cal T}(\Sigma^g)\equiv T^*({\cal A}^F/PSU(1,1))$ from (2.4),
given by

$$
\eqalign{
&\{ T^0(\a),T^0(\b)\}=0\cr
&\{ T^0(\a),T^1(\b)\}=-\frac12\,\sum_n\Delta_n(\a,\b)\,
\Big(T^0(\a\circ_n\b)-T^0(\a\circ_n\b^{-1})\Big)\cr
&\{ T^1(\a),T^1(\b)\}=-\frac12\,\sum_n\Delta_n(\a,\b)\,
\Big(T^1(\a\circ_n\b)-T^1(\a\circ_n\b^{-1})\Big),}\eqno(2.9)
$$

\ni where the sums are over all intersection points $n$ of
the homotopy elements $\a$ and $\b$, with $\Delta_n(\a,\b)=1$ ($=-1$)
if the two tangent vectors $(\dot\a,\dot\b)$ form a right-
(left-)handed zweibein at $n$. A similar algebra is derived by Nelson
and Regge in [16]. The algebraic Mandelstam constraints mentioned
earlier take the form

$$
\eqalign{
&T^0(\a)
T^0(\b)=\frac12\Big(T^0(\a\circ\b)+T^0(\a\circ\b^{-1})\Big)\cr
&T^0(\a) T^1(\b)+T^0(\b)
T^1(\a)=\frac12\Big(T^1(\a\circ\b)+T^1(\a\circ\b^{-1})\Big) }
\eqno(2.10)
$$

\ni for pairs of intersecting homotopy elements $\a$ and $\b$. Before
discussing quantum theories based on the algebraic relations (2.9)
and (2.10), we will present an even simpler quantization, defined
directly on the reduced phase space $T^*{\cal T}(\Sigma^g)$, the
cotangent bundle over Teichm\"uller space. Since Teichm\"uller space
is diffeomorphic to the real space $\R^{6g-6}$, one can simply
choose a set of global coordinates $x_i$, $i=1\dots 6g-6$, and a
corresponding set of momenta $p_i$, and quantize \`a la
Schr\"odinger on $L^2(\R^{6g-6},dx)$. This is straightforward, but
not terribly helpful as long as one does not establish an explicit
relation between the coordinates $x$ and the observables introduced
earlier. At this point we need a mathematical result by Okai [17],
who established an explicit cross section of the bundle of
$PSU(1,1)$-valued holonomies, the space of homomorphisms
Hom$(\pi_1(\Sigma^g),PSU(1,1))$ over Teichm\"uller space. As
coordinates on ${\cal T}(\Sigma^g)$ he uses the Fenchel-Nielsen
parameters [18], a set $(l_i,\tau_i)$, $i=1\dots 3g-3$, of length and
angle coordinates associated with a pants decomposition of the
genus-$g$ surface. Using this result, one may write arbitrary Wilson
loops as functions $T(\g)[l,\tau]$ of the Fenchel-Nielsen
coordinates, and thus make contact with the previous formulation.
It also enables one to find an explicit set of independent Wilson
loop coordinates on Teichm\"uller space, i.e. to solve the
overcompleteness problem [19].

If one regards this description of the reduced phase space as the
basic one, and the Wilson loop variables as derived quantities, one may
still want to represent the Wilson loops as well-defined operators
on the Hilbert space $L^2({\cal T}(\Sigma^g),dl\, d\tau)$, and
realize some or all of the algebraic relations (2.9), (2.10) in the
quantum theory. Alternatively, one may take the Wilson loop
observables as fundamental physical quantities, and try to find
self-adjoint representations of the $T^0$ and $T^1$, such that their
commutator algebra is isomorphic to the classical Poisson brackets
(2.9). A priori one expects this latter quantization ansatz to
be more general, since it involves the entire representation
theory (not just that on $L^2({\cal T}(\Sigma^g),dl\, d\tau)$) of a
rather complicated Poisson algebra, whereas the former is expected to
be essentially unique, up to possible factor orderings for the $\hat
T^0$ and $\hat T^1$ in terms of the fundamental operators $\hat l$,
$\hat\tau$, $\hat p_l$ and $\hat p_\tau$.

A similar issue arises in 3+1 quantum general relativity, where it
has been suggested to use a quantization based on $SL(2,\C)$-Wilson
loops [15]. The variables in this case are of course field-theoretic
and the analogue of the loop algebra (2.9) requires a
regularization. The question is whether one should abstractly study
the representation theory of this loop algebra or consider only
special representations that can be obtained (formally) through an
integral transform from the connection representation, where
$SL(2,\C)$-connections modulo gauge $A\in{\cal A}/{\cal G}$ are
regarded as fundamental. This so-called loop transform [15] has the
form

$$
\psi (\g):=\int_{{\cal A}/\cal G}d\mu(A) \ \
T^0(\g)[A]\ \Psi (A),\eqno(2.11)
$$

\ni where the Wilson loop functionals $T^0$ plays the role of an
integral kernel, and wave functions $\psi (\g)$ in the loop
representation are labelled by spatial closed curves $\g$. The idea
is that once the loop transform has been defined rigorously, one
obtains a loop representation that is unitarily equivalent to the
connection representation. In contrast with 3+1 dimensions, in 2+1
dimensions this construction can be carried out explicitly. This is
particularly useful since in practice it turns out to be
difficult to abstractly construct irreducible representations of
(2.9), with the operators simultaneously satisfying quantum analogues
of the constraints (2.10), constraints arising from (2.6), and other
conditions like $T^0\geq 1$.

An early implementation of these ideas can be found in a series
of papers [3], where a rigorous quantum loop representation for
$\Sigma=T^2$ is constructed, however, only the compact sector (where
all Wilson loops are bounded by $|T^0(\g)|\leq 1$) is quantized.
As discussed in detail by Marolf [20], this construction of the loop
representation via the loop transform cannot be carried over
unmodified to the physical, non-compact sector. The reason for this
is readily illustrated by the explicit form of the transform for the
torus-case (where the analogue of the Teichm\"uller space is
$\R^2/\Z_2$, parametrized by $a_1$, $a_2$), given by

$$
\psi(\vec n)=\;<T^0(\vec n),\Psi>\,=
\int d\vec a\; T^0(\vec n)\;\Psi(\vec a).\eqno(2.12)
$$

\ni Since the homotopy group $\pi_1 (T^2)$ is abelian, its elements
can be labelled by a pair $\vec n$ of integers. Since $T^0(\vec
n)=\cosh \vec n\cdot\vec a$, the integrand of (2.12) diverges rapidly
for large $\vec a$. As demonstrated in [20], the kernel of the
transform, i.e. those elements mapped to 0, is in fact dense in
$L^2(\R^2,d\vec a)$. Nevertheless, one may define loop
representations that are isomorphic to the connection
representation. This involves the choice of a dense subspace of
$L^2(\R^2,d\vec a)$, satisfying a number of properties, and as a
result of the construction in general contains wave functions that
cannot be expressed as functions of homotopy classes. Strictly
speaking, these are therefore not ``loop representations" in the
usual sense.

An alternative way of making the loop transform (2.12) well-defined
was proposed by Ashtekar and Loll [21]. The basic idea is to employ a
non-trivial volume element $dV=e^{-M(\vec a)} d\vec a$ in the
transform that provides a sufficient damping for large $\vec a$, so
as to make it converge for general elements of the connection
Hilbert space. That this is a viable procedure was demonstrated
in [21], where the loop representation on $T^2$ was constructed for a
particular choice of the damping factor $M(\vec a)$. The choice of a
suitable measure is an additional input, and $M$ has to
satisfy a number of conditions in order to make the loop
representation well-defined. The explicit form of the
$\hat T^1$-operators and their action on loop states $\psi (\vec n)$
is more complicated than in the case for trivial measure, since it
contains a contribution from $\nabla M$.
Still, by construction this loop representation
is isomorphic to the connection representation and, in particular,
quantum analogues of (2.9) and (2.10) continue to hold.

Of these two approaches, only the Ashtekar-Loll construction has been
extended to the higher-genus case, although not in as much detail
as in the torus case. In a first step, let us point out that within
the connection representation on $L^2({\cal T}(\Sigma^g),dl\,
d\tau)$, one can straightforwardly construct self-adjoint operators
$\hat T^0$ and $\hat T^1$ corresponding to (2.8). Using the results
of [17], one can write any Wilson loop $T^0$ as a function of the
Fenchel-Nielsen coordinates $l_i$, $\tau_i$. The corresponding
self-adjoint operators act as multiplication operators. To find the
momentum operators $\hat T^1$, one uses the fact that there is a
natural symplectic structure on Teichm\"uller space (although ${\cal
T} (\Sigma)$ presently plays the role of a {\it configuration}
space), namely the Weil-Petersson symplectic form $\sum_i dl_i\wedge
d\tau_i$, with respect to which the Fenchel-Nielsen coordinates are
canonical. As was already discussed in [3], this structure can be
used to obtain an explicit representation for the momentum operators
$\hat T^1$. As an example, consider the Wilson loops of the pair of
homotopy generators $\a_1$, $\b_1$ for $g=2$ as functions of the six
Fenchel-Nielsen coordinates $l_{-\infty}$, $l_0$, $l_\infty$,
$\tau_{-\infty}$, $\tau_0$, $\tau_\infty$ (see [17,19] for
derivation and notation),

$$
\eqalign{
\hat T^0(\alpha_1) &= \cosh{\frac{l_{-\infty}}{2}}\cr
\hat T^0(\beta_1) &=
\sinh{\frac{\tau_{-\infty}}{2}}\sinh{\frac{\tau_0}{2}}
+\frac{\cosh{\frac{l_{-\infty}}{2}}\cosh{\frac{l_0}{2}} +
\cosh{\frac{l_\infty}{2}}}{\sinh{\frac{l_{-\infty}}{2}}
\sinh{\frac{l_0}{2}}}
\cosh{\frac{\tau_{-\infty}}{2}}\cosh
{\frac{\tau_0}{2}}.}\eqno(2.13)
$$

\ni The corresponding self-adjoint momentum operators are

$$
\eqalign{
\hat T^1(\a_1) &=-\frac{i\hbar}{2}\sinh{\frac{l_{-\infty}}{2}}
\frac{\del}{\del\tau_{-\infty}}\cr
\hat T^1(\b_1) &=-\frac{i\hbar}{2}
\frac{\cosh{\frac{l_{\infty}}{2}}\cosh{\frac{l_0}{2}} +
\cosh{\frac{l_{-\infty}}{2}}}{\sinh^2{\frac{l_{-\infty}}{2}}
\sinh{\frac{l_0}{2}}}
\cosh{\frac{\tau_{-\infty}}{2}}\cosh{\frac{\tau_0}{2}}
\;\frac{\del}{\del\tau_{-\infty}}\cr
-&\frac{i\hbar}{2}
\frac{\cosh{\frac{l_{\infty}}{2}}\cosh{\frac{l_{-\infty}}{2}} +
\cosh{\frac{l_0}{2}}}{\sinh{\frac{l_{-\infty}}{2}}
\sinh^2{\frac{l_0}{2}}}
\cosh{\frac{\tau_{-\infty}}{2}}\cosh{\frac{\tau_0}{2}}
\;\frac{\del}{\del\tau_0}\cr
+&\frac{i\hbar}{2} \frac{\cosh{\frac{\tau_{-\infty}}{2}}
\cosh{\frac{\tau_0}{2}}}{\sinh{\frac{l_{-\infty}}{2}}
\sinh{\frac{l_0}{2}}}\sinh{\frac{l_\infty}{2}}
\;\frac{\del}{\del\tau_\infty}\cr
-&\frac{i\hbar}{2}\big( \cosh{\frac{\tau_{-\infty}}{2}}
\sinh{\frac{\tau_0}{2}}+
\frac{\cosh{\frac{l_{-\infty}}{2}}\cosh{\frac{l_0}{2}} +
\cosh{\frac{l_\infty}{2}}}{\sinh{\frac{l_{-\infty}}{2}}
\sinh{\frac{l_0}{2}}} \sinh{\frac{\tau_{-\infty}}{2}}
\cosh{\frac{\tau_0}{2}} \big) \;\frac{\del}{\del l_{-\infty}}\cr
-&\frac{i\hbar}{2}\big( \cosh{\frac{\tau_0}{2}}
\sinh{\frac{\tau_{-\infty}}{2}}+
\frac{\cosh{\frac{l_{-\infty}}{2}}\cosh{\frac{l_0}{2}} +
\cosh{\frac{l_\infty}{2}}}{\sinh{\frac{l_{-\infty}}{2}}
\sinh{\frac{l_0}{2}}} \sinh{\frac{\tau_0}{2}}
\cosh{\frac{\tau_{-\infty}}{2}} \big) \;\frac{\del}{\del l_0},}
\eqno(2.14)
$$

\ni where we have chosen a factor ordering with the momenta to the
right. The functional form of the Wilson loop operators is
considerably more complicated than the corresponding expressions in
the torus case. In [19] it is shown that also for the higher-genus
case there exist suitable measures that ensure the convergence of
the loop transform for a sufficiently big set of connection wave
functions. Thus there are no obvious obstacles to quantizing along
the lines proposed in [21], although the details of these loop
representations remain to be worked out.

There is a different treatment by Manojlovi\`c and Mikovi\`c in the
connection formulation [22], which is not based on the classical
reduction to the reduced phase space, but instead relies on a quantum
reduction \`a la Dirac. For a non-vanishing spatial determinant
${}^{(2)}g$, one may rewrite the action (2.3) in such a way that the
functional form of the ensuing first-class constraints is exactly
analogous to the ones  obtained in the Ashtekar formulation in 3+1
dimensions [23]. In particular, in this form the Hamiltonian
constraint is quadratic in the momenta $E$. For the torus case, one
obtains an effective finite-dimensional theory with three Gauss law
constraints and one Hamiltonian constraint. It is argued that the
quantum theory is given by unitary irreducible representations with
zero mass of the Poincar\'e algebra in three dimensions. Since the
states in these representations depend on two real parameters, this
suggests that the reduced configuration space of the system (2.3) is
$\R^2$, which does not quite agree with the usual result.
Probably this can
be traced to a subtlety in the solution to the Gauss law
constraints, which may be given in terms of wave functions of three
rotationally invariant parameters $a_i$. These are treated as free
parameters in [22], whereas strictly speaking they are subject to a
number of inequalities ($|a_3|\leq a_1a_2$, $a_1\geq 0$, $a_2\geq
0$). \vskip1cm
\line{\titit 2.2 Non-vanishing cosmological constant\hfil}

Let us now turn to the cases with a non-vanishing cosmological
constant. As discussed earlier, their physical phase spaces too are
given by spaces of flat connections modulo gauge. One may
therefore again describe them as
suitably regular spaces of homomorphisms of
$\pi_1(\Sigma^g)$ into the gauge groups $G=SO(2,2)$ and $SO(3,1)$,
for $\Lambda <0$ and $\Lambda >0$ respectively. However, since those
groups do not have a cotangent bundle structure, the holonomies
and Wilson loop variables are now necessarily functions on phase
space (unlike the Wilson loops $T^0$ of (2.8), that are functions on
configuration space).

In [24], Nelson, Regge and Zertuche compute the
path-dependent Poisson algebras for the $G$-valued phase space
holonomies and,
after going to the spinor representations $SL(2,\R)\times SL(2,\R)$
and $SL(2,\C)$ respectively, the Poisson algebra of the corresponding
Wilson loops $T(\g)[A]=\frac12 {\rm Tr}\ U_\g$.
In the former case, one gets two copies $\{ T^+(\g)\}$ and $\{
T^-(\g)\}$ of $SL(2,\R)$-Wilson loops satisfying the Poisson algebra
(c.f. (2.9))

$$
\eqalign{
&\{ T^\pm (\a),T^\pm (\b)\}=\pm \frac{\Delta}{4}\sqrt{-\Lambda}\,
\big(T^\pm (\a\circ\b)-T^\pm (\a\circ\b^{-1})\big)\cr
&\{T^+ (\a),T^- (\b)\}=0,}\eqno(2.15)
$$

\ni for pairs of homotopy elements $\a$, $\b$ with a single
intersection. For $\Lambda >0$, a similar decomposition is only
possible over the complex numbers, and the factor on the
non-vanishing right-hand side in (2.15) has to be replaced by
$i\frac{\Delta}{4}\sqrt{\Lambda}$, which is purely imaginary.
They go on to study the representation theory of the
``plus-sector" $\{ T^+(\g)\}$ of the algebra (2.15), restricted to a
single ``handle", i.e. to the subgroup of $\pi_1(\Sigma^g)$
generated by a single pair of generators $\a_i$ and $\b_i$. The
motivation for this is the hope that the full quantum theory of a
genus-$g$ surface may be obtained by combining several such copies
appropriately, although a concrete construction to our knowledge has
not yet been given.  In any case, the quantization for one handle
they propose is based on a (non-Lie) algebra of a rescaled version
of the operators $\hat T^+(\a)$, $\hat T^+(\b)$ and $\hat
T^+(\a\circ\b)$. The physically relevant quantum representations are
those where the basic operators are unbounded, and [24] contains a
preliminary discussion of some of their properties.

This investigation for $\Lambda <0$ is extended to the case $g=2$ by
Nelson and Regge in [25-28], where it is proposed
to base the quantization on
a ring $\cal R$ of polynomials of a highly symmetric subset of 15
Wilson loop variables $\{ T^+(\g)\}$. $\cal R$ is closed under
Poisson brackets and the subset is chosen so that any other traced
holonomy may be expressed as a function of this subset via the
Mandelstam identities (the first set of relations in (2.10)). To
eliminate the remaining
overcompleteness of these observables in the classical
theory, they propose a quotient construction in which the physical
observables are elements of ${\cal R}/I({\cal R})$, where $I({\cal
R})$ is an ideal (closed under Poisson brackets), generated by the
Mandelstam constraints (here called ``rank identities") and
constraints coming from the fundamental relation (2.6), called
``trace identities". The difficulty lies in finding an appropriate
basis for this ideal, which should consist of 6 rank and 3 trace
identities. The ideals for $g=1,2$ are described in [27], where it is
also argued that a similar quotient space construction should be
applied to the higher-genus case. (Note that the results in [19]
may be used to explicitly parametrize the quotient spaces ${\cal
R}/I({\cal R})$.)

An issue we have not touched upon so far is the role of the large
diffeomorphisms $\frac{{\rm Diff}\,\Sigma}{{\rm Diff}_0 \Sigma}$, i.e.
those that do not lie in the component ${\rm Diff}_0 \Sigma$ connected
to the identity. They form the so-called mapping class group, also
called the Teichm\"uller modular group, whose generators are the
Dehn twists. The question is whether or
not one should regard them as gauge degrees of freedom, to be
factored out like the connected diffeomorphisms. The canonical Dirac
treatment of constraints only requires invariance under the action
of the connected component of a gauge group. For 2+1 gravity, there
is a whole spectrum of proposals how the large diffeomorphisms
should be treated in both the classical and the quantum theory, that
goes from ignoring them altogether over implementing them as unitary
symmetries to requiring strict invariance, even in the quantum
theory (see also the discussion in [29]). In principle such a
controversy should be settled by physical arguments, but this
presents a problem for a theory like three-dimensional gravity that
is largely unphysical. We therefore do not expect that this issue
has a definite resolution, and what remains to be understood is
which approaches to the large diffeomorphisms are feasible in
practice.

Note that none of the Wilson loop variables introduced so far are
invariant with respect to large diffeomorphisms. However, they
carry (more or less complicated) actions of the mapping class group.
For $g=2$, Nelson and Regge investigate a canonical (non-linear)
action of the Dehn twists on
the algebra ${\cal R}$ introduced above [25]. They also study the
centre of this algebra with respect to the Dehn twists, i.e. its
invariant elements. For genus $g$, they find $g+1$ such central
elements, out of which two remain linearly independent once the rank
identities are taken into account [27]. However, it is not explained
whether or how this construction intertwines with the quotient
construction of ${\cal R}/I({\cal R})$ to yield classical
observables that are
invariant under the mapping class group.

It turns out to be rather non-trivial to quantize the algebraic
structure of the classical algebra ${\cal R}$ of Wilson loop
variables, and the quotient construction for the physical
observables. A quantum analogue of the classical Poisson algebra of
the polynomials in the 15 chosen $T^+$'s is given in [26]. Since the
algebra is polynomial, this involves a particular choice of operator
ordering. The commutation relations involve a complex constant $K$ that
depends on the cosmological constant $\Lambda$ and goes to $1$ as
$\hbar\rightarrow 0$. In this limit, the classical Poisson brackets
are recovered by substituting

$$
\frac{1}{K-1}\, [\hat T^+(\g_1),\hat T^+(\g_2)]\rightarrow
\{ T^+(\g_1),T^+(\g_2) \}.\eqno(2.16)
$$

\ni Similarly, a $K$-dependent quantum action of the Dehn twists on
the quantized algebra $\hat {\cal R}$ can be defined. This framework
seems suggestive of a quantum theory defined on a Hilbert space
$L^2(\R^{15})$, where the quantum counterparts of the classical
constraints remain to be imposed to project out the physical wave
functions. However, it seems to be difficult to carry out this
program explicitly, as well as to implement modular invariance at the
quantum level. These observations are in line with remarks made
earlier in the context of the ($\Lambda =0$)--case. It therefore
may not come as a total surprise that in their most recent paper
on the $g=2$, $\Lambda <0$ quantum gravity, Nelson/Regge propose a
quantization based on a reduced
set of 6 variables, three angles $\varphi_a$
and conjugate momenta $p_a$, $a=1\dots 3$, in terms of which all of
the 15 Wilson loop variables can be expressed [28]. (Recall that the
dimension of the physical phase space for $g=2$ is 12, and that we
have split it into two $SL(2,\R)$-sectors.) The quantum operators
$\hat T^+(\a)$ are functions of the basic operators $\hat\varphi_a$
and $\hat p_a$, and depend on a complex parameter $\Theta$, where
$e^{i\Theta} =K$ (and the classical limit therefore corresponds to
$\Theta =0$). For example, the form one finds for the classical
Wilson loops of the fundamental homotopy generators $\a_1$ and
$\b_1$ is

$$
\eqalign{
&T^+(\a_1)=\frac{\cos \varphi_2}{\cos\frac{\Theta}{2}}\cr
&T^+(\b_1)=\frac{1}{2\cos\frac{\Theta}{2}}\sum_{n,m=\pm1}
\frac{\sin (\frac{\Theta}{4}+\frac{n\varphi_1
+m\varphi_2+\varphi_3}{2} )
      \sin (\frac{\Theta}{4}+\frac{n\varphi_1
+m\varphi_2-\varphi_3}{2} )}
     {\sin n\varphi_1\;\sin m\varphi_2}\times \cr
&\hskip6cm e^{-i (\frac32 n\varphi_1 +
      \frac32 m\varphi_2+\Theta (n p_1+m p_2))}.}\eqno(2.17)
$$

\ni One observes that, in contrast with (2.14), the conjugate
momentum operators $\hat p_a=-i\frac{\del}{\del\varphi_a}$
will not appear
linearly, but exponentially in the corresponding quantum operator
$\hat T^+(\b_1)$. In this ``reduced phase space quantization", the
trace and rank identities are fulfilled both classically
and quantum-mechanically. What is slightly puzzling about this
approach is the fact that one seems to end up with three free
canonical coordinate pairs, although one knows that the underlying
moduli space of $SO(2,2)$- (or $SL(2,\R)\times SL(2,\R)$-)connections
is not a cotangent bundle.

The Hilbert space proposed in [28] is an $L^2$-space on a
suitable domain $D^3\subset \R^3$ based on three real parameters
$z_a=\cos\varphi_a$. The inner product on this Hilbert space is to be
determined by requiring the basic operators to be self-adjoint. The
physically interesting case is the one where the angles $\varphi_a$ are
imaginary and therefore $z_a\geq 1$. An appropriate scalar product
for this case remains to be found.

Another quantization method for $\Lambda <0$
has been suggested by Ezawa [14], who
aims at constructing a unified quantization for all sectors of the
reduced Chern-Simons moduli space, not just that corresponding to the
geometrodynamic solutions. He adopts a ``brute-force" approach
(i.e. without any physical justification) to
make this space into a cotangent bundle, to which then a geometric
quantization procedure (in the sense of Kostant and Souriau) may be
applied.
\vskip1cm
\line{\titit 2.3 Summary\hfil}

This concludes our discussion of quantizations in the gauge-theoretic
approach to 2+1 gravity.
The most promising approaches seem to be those that are closest to
a Schr\"odinger-type quantization on the reduced physical phase
space. For $\Lambda =0$, the construction of a well-defined quantum
theory is straightforward, and in addition one can show that the
physically interesting Wilson loop observables can be defined as
self-adjoint operators in this representation. The existence of
appropriate transforms ensures that there are well-defined
loop representations, whose wave functions are labelled by homotopy
classes. This shows that the non-compactness of the gauge group does
not present any problems in principle to constructing such
representations. However, since they are by construction unitarily
equivalent to the reduced phase space quantization, they do not
yield a priori any additional physical information.

The situation for $\Lambda\not= 0$ is not as straightforward, since
the lack of a cotangent bundle structure prevents an analogous
construction of the reduced phase space quantization.
Given such a quantum theory, for example, as the result of a
geometric quantization based on a complex
polarization on the reduced phase space [1,10], one could again
attempt to define the Wilson loop operators on its Hilbert space.
As we have seen, the existing quantization proposals for
$\Lambda\not= 0$ are based on algebras of Wilson loops, and the
details beyond genus-1 become rather involved. Furthermore,
it is in general difficult to find quantum analogues of the
constraints satisfied by the classical Wilson loops. Although the
most recent proposal of Nelson and Regge for $g=2$, $\Lambda <0$ is
based on wave functions of three variables (and therefore looks
like a ``reduced quantization") [28], it is not clear how these
relate to an explicit parametrization of the reduced phase space.

Incorporating invariance under the mapping class group seems
problematic in all of the connection approaches, because the natural
classical observables, the Wilson loops, are not modular invariant.
If we regard the large diffeomorphisms as gauge degrees of freedom,
the only true classical observables are the Casimir operators of
Nelson/Regge, which are far from forming a complete set. This has to
do with the fact that the action of the Dehn twists on the reduced
phase spaces is rather complicated, so that even in the
torus-case there is no simply defined ``fundamental region" for the
modular action. As discussed by Peld\`an [29], this also leads to
problems if one tries to find finite-dimensional representations of
the modular group on connection wave functions (invariance being
just a special case).

\vskip1.5cm
\line{\ch 3 Quantization in the geometric formulation\hfil}

The classical starting point for a quantization within the geometric
formulation is the Lagrangian (1.1). Also in this description one
can reduce the degrees of freedom to a finite number, as was already
noted by Martinec in 1984 [30], and later rediscovered by Hosoya and
Nakao [31] and Moncrief [32]. We will discuss in the following the
case where $\Lambda =0$, although most of the classical treatment is
readily extended to include a cosmological constant [33,11]. For
general genus, the three constraints ${\cal H}=0$ and ${\cal H}_i=0$
can be decoupled and subsequently solved if one adopts the York time
gauge, where $\Sigma^g$ is taken to be a constant mean curvature
surface [31,32], i.e. the ``time" parameter $\tau$ is given by

$$
\tau = \frac{1}{\sqrt{{}^{(2)} g}}\, g_{ij}\pi^{ij}.\eqno(3.1)
$$

\ni For $g>1$, the two-metric $\g_{ij}$ can be uniquely decomposed into
a conformal factor $e^{2\lambda (x)}$ and a metric $h_{ij}$ of constant
scalar curvature $-1$. The space ${\cal M}_{-1}$ of such
metrics is infinite-dimensional, but one may show that the quotient
by the diffeomorphisms, ${\cal M}_{-1}/{\rm Diff}_0$ is well-defined
and diffeomorphic to the Teichm\"uller space ${\cal T}(\Sigma^g)$
[32]. The physical phase space is given by the cotangent bundle
$T^*{\cal T}(\Sigma^g)$, with global coordinates $(m_a,p^a)$,
$a=1\dots 6g-6$. The Hamiltonian constraint ${\cal H}=0$ determines the
conformal factor $\lambda$ as a $\tau$-dependent function on this
cotangent bundle. The action (1.1) in the reduced variables becomes

$$
\int d\tau\; \big( p^a \frac{dm_a}{d\tau}-H(m,p,\tau)\big),
\eqno(3.2)
$$

\ni where $H=\int_{\Sigma^g} d^2x \sqrt{{}^{(2)} g}\
e^{2\lambda (m,p,\tau)}$ is the Hamiltonian associated with the
York time slicing. This
Hamiltonian measures the area of the spatial surfaces of constant
$\tau$, and generates the time evolution on $T^*{\cal T}(\Sigma^g)$.
Unfortunately the solution for $\lambda$ is known only implicitly, as
a solution of a differential equation [32], which is problematic
since  $H$ depends on it explicitly. Consequently, the classical and
quantum theory have been studied in detail only for the torus case,
where the Hamiltonian $H$ is known as a function of the basic
variables. We will concentrate on this case in the following,
starting with $\Lambda =0$.
There are two canonical pairs with $\{m_a,p^b\}=\delta_a^b$ and the
Hamiltonian is given by

$$
H(m,p,\tau)=\frac{1}{\tau}\sqrt{m_2^2\big( (p^1)^2+(p^2)^2\big) }.
\eqno(3.3)
$$

\ni Alternatively, if one wants to get rid of the
square root, one may employ a different gauge condition. For example,
Martinec chooses the two-metric to be spatially constant,
$g_{ij}(x,t)=g_{ij}(t)$ [30], and Hosoya/Nakao the lapse function to
be spatially constant, $N(x,t)=N(t)$ [31]. This leads to a reduced
action of the form

$$
S=\int dt\;\big( p^a \frac{dm_a}{dt}+\tau\frac{dv}{dt}-N {\cal
H}(m,p,v,\tau)\big),\eqno(3.4)
$$

\ni with a Hamiltonian constraint $\cal H$ quadratic in both the
momenta $p^a$ and the variable $\tau$ canonically conjugate to the
volume variable $v$. Going to the quantum theory, in the first
approach one looks for solutions of the Schr\"odinger equation

$$
i\,\frac{\del\psi}{\del\tau}=\hat H\psi(m,\tau),\eqno(3.5)
$$

\ni whereas in the second one tries to solve the quantum Hamiltonian
constraint  $\hat {\cal H}\psi=0$, which takes the form of a
Klein-Gordon equation. The latter form is more convenient because of
the absence of the square root. As wave functions one may take
either the ``volume representation" on states $\psi(m,v)$, or the
``time representation" on $\psi(m,\tau)$. Martinec chooses the
former since he is also interested in the case with non-vanishing
cosmological constant (for which the Hamiltonian constraint contains
a term proportional to $\Lambda v$), and gives the general form of
the solution [30]. Hosoya and Nakao [34] impose the Hamiltonian
constraint on wave functions $\psi(m,s)$, where $s=\ln v$, but in
addition insist that physical states should be invariant under large
diffeomorphisms, which in the torus case are elements of the group
$SL(2,\Z)$. They propose to superimpose (non-invariant) solutions to
$\hat {\cal H}\psi=0$ in order to arrive at $SL(2,\Z)$-invariant
wave functions. However, this construction remains somewhat
implicit, because it involves Maass forms, that have been the subject
of much study, but are not all known explicitly as functions of the
modular parameters $m_a$. (The Maass forms are the modular-invariant
eigenfunctions of the Laplacian on Teichm\"uller space, that comes
from the quantization of the term under the square root in
(3.3). Its spectrum has a continuous and a discrete part and Puzio
has recently suggested that both should be taken into account when
constructing physical wave functions [35].)

These ideas are taken up by Carlip [36], who argues that (with an
appropriate operator ordering) the Schr\"odinger equation (3.5)
should be regarded as the positive square root of the Wheeler-DeWitt
equation $\hat {\cal H}\psi (m,\tau)=0$. (How one may rigorously make
sense of the square root operation at the quantum level is
discussed in [35].) He also attempts to define a Hilbert space, i.e.
an inner product on the solution space, an issue not considered by
previous authors. He proposes $\int \frac{d^2m}{m_2^2}$ as a
modular-invariant inner product, where the integration domain is
taken to be a fundamental region for the modular group in the upper
half plane for the complex variable $m:=m_1+i\ m_2$. The $\tau$ in
$\psi (m,\tau)$ is therefore considered as an external parameter.
With this scalar product, the momentum operators $\hat p^a=-i\
\frac{\del}{\del m_a}$ are not self-adjoint, although the Laplacian
appearing in the Hamiltonian is. This leads one to consider a
different representation for the basic operator $\hat p_2$, namely
$\hat p_2=-i\ \frac{\del}{\del m_2}+\frac{i}{m_2}$, which makes it
self-adjoint, provided the wave functions obey appropriate fall-off
conditions at the integration boundaries for $m_1$ and $m_2$.
However, it turns out that for consistency the wave functions $\psi$
then have to transform as forms of modular weight $\frac12$. This
construction is extended in [37] to representations on spaces of
weight functions of arbitrary modular weight, and leads to a
one-parameter family of quantum Hamiltonians, that according to
[37,12] give rise to inequivalent quantum theories.

The quantum equivalence between the reduced connection formulation
and the reduced geometric formulation for $\Sigma =T^2$ is considered
in [38]. The aim is to construct a quantum analogue of the
time-dependent canonical transformation between the corresponding
classical theories [38] (see also [39]). However, since the canonical
transformation contains inverses of the momenta, the corresponding
operators are only formally defined. Still, a formal operator
problem arises, and it is shown that this ambiguity is reduced when
demanding the quantum operators to have the same modular
transformation behaviour as their classical counterparts. (An
analogous procedure is used in [12] to relate the Nelson-Regge torus
quantum theory for $\Lambda <0$ with the ADM quantization.)  A
similar line of argument is followed by Anderson [40], who defines
another quantum canonical transformation between the reduced
holonomy and the geometric approach. This is defined rigorously, but
requires a peculiar ``operator-valued measure density" in the
definition of the scalar product for the connection representation,
that is different from $\int d\vec a$, but likewise modular
invariant. The author alludes to the fact that the use of such
measure densities leads to a new ambiguity in the quantization. It
would be interesting to understand the physical significance of this
generalized quantum structure.

For completeness, let us mention that there is a proposal
to construct modular-invariant wave functions in the connection
representation for the torus via an integral transform from the
geometric approach in terms of Maass form wave functions [36,37].
However, as has been pointed out elsewhere [13,29], this construction
is rather subtle and not well-defined in the form proposed there.
This presumably affects also the analogous quantum constructions for
$\Lambda \not= 0$ in [11].

Carlip considers yet another quantization approach, in the form of
a more complicated Wheeler-DeWitt equation, obtained
when no time-slicing is imposed classically, i.e. the function
$\lambda$ in the conformal factor $e^{2\lambda}$ is left arbitrary
[41]. As a result, the Wheeler-DeWitt equation is a non-local
equation on wave functions $\Psi (m,\lambda)$, containing functional
derivatives with respect to $\lambda(x)$, and too complicated to
be solved directly. To enable comparison with the gauge-fixed
ADM wave functions, he introduces a formal functional
Fourier transform $\Psi(m,\lambda)\mapsto\tilde\Psi(m,\tau)$, but the
resulting equation for $\tilde\Psi(m,\tau)$ is not in any obvious way
equivalent to the ones discussed in the fully reduced ADM
formulation. Attempting to gauge-fix \`a la Faddeev/Popov to obtain a
scalar product for wave functions in the York time gauge leads to a
highly complicated, operator-valued Faddeev-Popov determinant, whose
structure is known only  perturbatively.

A similar quantization is suggested by Visser in [42], who
also formulates the Wheeler-DeWitt equation on the superspace of the
infinite-dimensional space of conformal factors times the
finite-dimensional moduli space, but obtains an equation not
containing any non-local terms. He argues that the Hamiltonian
constraint splits into independent constraints on the conformal mode
and the modular parameters, and that the latter should contain,
besides the Laplacian, also a term proportional to the Ricci scalar
on moduli space, however, no solutions are given. -- This ends our
overview of the geometric formulations of 2+1 quantum gravity.

\vskip1.5cm
\line{\ch 4 Conclusions\hfil}

Let us try to draw some conclusions from the quantization attempts
described above. From the point of view of the generic higher-genus
case, the quantization on the reduced connection phase space -- at
least for $\Lambda =0$ -- is the one furthest developed and
potentially most promising, and allows the Wilson loop observables
to be defined as well-defined quantum operators. On the other hand,
looking for abstract representations of algebras of Wilson loops not
based on a Schr\"odinger-type quantization of the reduced
phase space seems to be much harder. For the case of $\Lambda \not= 0$,
further illumination is needed of how the non-cotangent bundle
structure of the moduli spaces is reflected in the quantum
theory. In fact, the case $\Lambda >0$ has hardly been explored
(see, however, the discussion in [10], which contains some
suggestions of how the gravitational Hilbert space may be related to
that of a Chern-Simons theory with complex gauge group).

If one is not content with such a quantization of ``frozen
dynamics" in terms of time-independent constants of motion, one has
to consider an ADM-type quantization, which at this moment does not
seem feasible beyond $g=1$. In this approach -- although a priori
undesirable -- one is in practice restricted to a particular choice
of gauge-fixing, the York time gauge. Quantization of this gauge
degree of freedom poses difficulties that so far have not been
overcome. In the ADM treatment of the torus case, in contrast with
the connection formulation, incorporation of modular invariance does
not present any obvious problems, although the resulting Hilbert
spaces are not known in a very explicit way.

Since in the best-explored case of $\Sigma=T^2$, the relation between
the classical formulations in terms of Teichm\"uller parameters and
holonomies, and reduced ADM variables is well-understood [38,39], it
is natural to search for a corresponding relation in the quantum
theory.  It is relatively easy to establish a formal correspondence
between operators in the various quantizations (by ``putting hats on
everything"), but it seems difficult to make these constructions
rigorous. This is not particularly surprising, and in essence a
consequence of the Groenewold-Van Hove theorem. If one has a quantum
theory in which a complete set of basic operators is represented by
self-adjoint operators, it is in general {\it not} possible to
represent another  quantity, that classically is a non-polynomial
function of those basic variables, as a self-adjoint operator
on the same Hilbert space. This
makes it hard to relate the quantum theories of the metric and the
connection approaches.

This may seem an unsatisfactory state of affairs, but on the other
hand it is well-known that not every classical equivalence can be
elevated to a quantum equivalence. Moreover, for phase spaces not
of the form of an $\R^{2n}$, the quantization is typically
non-unique, even if one starts from a single classical
description. For more physical theories, one may of course decide
that one quantum theory rather than another is correct, because it
is in better agreement with physical observations, but this road is
not available to us in the case of 2+1 gravity.

It has been suggested to resolve the ambiguity in the
choice of a time slicing in the ADM quantum theory
by declaring the time-less connection quantization as
fundamental [12], which is an interesting idea. From what we
have said above it follows that (for $g=1$) one should expect to
recover the ADM quantum theory at most in some appropriate
perturbative or semi-classical sense. In turn, one may in the same
limit try to {\it define} a (perturbative) quantum theory in the
geometric formulation for $g\geq 2$ via a connection quantization.

When attempting to generalize any of the above conclusions to 3+1
canonical gravity, one
should keep in mind that its structural resemblance is
greatest with that of 2+1 gravity for $g\geq 2$, as explained by
Moncrief [43]. Recall that also in 3+1 dimensions one has the choice
between a geometric formulation in terms of the four-metric
$g_{\mu\nu}$ and a connection formulation in terms of the Ashtekar
connection $A_\mu^a$, and that also in this case the quantization of
the latter has progressed much further than that of the traditional
ADM approach. One may therefore again be tempted to regard this
approach as fundamental as far as the quantum theory is concerned.
However, note that in this case the Wilson loop operators solve the
spatial diffeomorphism and Gauss law constraints, but not the quantum
Hamiltonian, and therefore the ``problem of time" cannot be solved
in the same way as suggested for the 2+1 theory.

\vskip2cm
\vfill\eject
\line{\ch References\hfil}

\item{[1]} E. Witten: 2+1 dimensional gravity as an exactly soluble
  system, {\it Nucl. Phys.} B311 (1989) 46-78
\item{[2]} I. Bengtsson: Yang-Mills theory and general relativity in
  three and four dimensions, {\it Phys. Lett.} B220 (1989) 51-4
\item{[3]} A. Ashtekar, V. Husain, C. Rovelli, J. Samuel, L. Smolin:
  2+1 gravity as a toy model for the 3+1 theory, {\it Class. Quant.
  Grav.} 6 (1989) L185-193; L. Smolin: Loop representation for
  quantum gravity in 2+1 dimensions, in: {\it Knots, topology and
  quantum field theory}, Proceedings of the 12th Johns Hopkins
  Workshop; A. Ashtekar: Lessons from 2+1 dimensional quantum gravity,
  in: {\it Strings 90}, ed. R. Arnowitt et al., World Scientific,
  Singapore
\item{[4]} A. Ashtekar and J. Romano: Chern-Simons and Palatini
  actions and 2+1 gravity, {\it Phys. Lett.} B229 (1989) 56-60
\item{[5]} S. Carlip: Six ways to quantize (2+1)-dimensional
  gravity, gr-qc/9305020, to appear in {\it Proc. of the 5th Canadian
  Conference on General Relativity and Relativistic Astrophysics}
\item{[6]} R. Loll: Chromodynamics and gravity as theories on loop
  space, hep-th/9309056, to appear in {\it Memorial Volume for M.K.
  Polivanov}
\item{[7]} J.N. Goldberg, J. Lewandowski, and C. Stornaiolo:
  Degeneracy in loop variables,{\it Comm. Math. Phys.} 148 (1992)
  377-402; A. Ashtekar and J. Lewandowski:
  Completeness of Wilson loop functionals on the moduli space of
  $SL(2,\C)$ and $SU(1,1)$
  connections, {\it Class. Quant. Grav.} (1993) L69-74.
\item{[8]} R. Loll: Loop variable inequalities in gravity and
  gauge theory, {\it Class. Quant. Grav.} 10 (1993) 1471-6
\item{[9]} G. Mess: Lorentz spacetimes of
  constant curvature, preprint Institut des Hautes \'Etudes Scientifiques
  IHES/M/90/28
\item{[10]} E. Witten: Quantization of Chern-Simons gauge theory with
  complex gauge group, {\it Comm. Math. Phys.} 137 (1991) 29-66
\item{[11]} K. Ezawa: Classical and quantum evolutions of the de
  Sitter
  and the anti-de Sitter universes in 2+1 dimensions, {\it Phys. Rev.}
  D49 (1994) 5211-26; Add.: ibid., D50 (1994) 2935-8
\item{[12]} S. Carlip and J.E. Nelson: Comparative quantizations of
  (2+1)-dimensional gravity, preprint Davis UCD-94-37 and Torino
  DFTT/49/94, gr-qc/9411031
\item{[13]} J. Louko and D. Marolf: The solution space of 2+1
  gravity on $\R\times T^2$ in Witten's connection formulation,
  {\it Class. Quant. Grav.} 11 (1994) 311-30
\item{[14]} K. Ezawa: Chern-Simons quantization of (2+1)-Anti-De
  Sitter gravity on a torus, pre\-print Osaka OU-HET/201,
  hep-th/9409074
\item{[15]} C. Rovelli and L. Smolin: Loop space representation of
  quantum general relativity, {\it Nucl. Phys.} B331 (1990) 80-152
\item{[16]} J.E. Nelson and T. Regge: Homotopy groups and 2+1
  dimensional quantum gravity, {\it Nucl. Phys.} B328 (1989)
  190-202
\item{[17]} T. Okai: An explicit description of the Teichm\"uller
  space as  holonomy representations and its applications, Hiroshima
  Math. J. 22 (1992) 259-71
\item{[18]} W. Abikoff: {\it The real analytic theory of Teichm\"uller
  space}, Lecture Notes in Mathematics, vol. 820, Springer, Berlin, 1980
\item{[19]} R. Loll: Independent loop invariants for 2+1 gravity,
  preprint Penn State Univ. CGPG-94/7-1, gr-qc/9408007
\item{[20]} D.M. Marolf: Loop representations for 2+1 gravity on a
  torus, {\it Class. Quant. Grav.} 10 (1993) 2625-47
\item{[21]} A. Ashtekar and R. Loll: New loop representations for
  2+1 gravity, {\it Class. Quant. Grav.} 11 (1994) 2417-34
\item{[22]} N. Manojlovi\'c and A. Mikovi\'c: Ashtekar formulation
  of (2+1)-dimensional gravity on a torus, {\it Nucl. Phys.} B385
  (1992) 571-86
\item{[23]} A. Ashtekar: {\it Lectures on Non-Perturbative Canonical
  Gravity}, World Scientific, Singapore, 1991
\item{[24]} J.E. Nelson, T. Regge, F. Zertuche: Homotopy groups and
  (2+1)-dimensional quantum De Sitter gravity, {\it Nucl. Phys.}
  B339 (1990) 516-32
\item{[25]} J.E. Nelson and T. Regge: 2+1 gravity for genus $>1$,
 {\it Comm. Math. Phys.} 141 (1991) 211-23
\item{[26]} J.E. Nelson and T. Regge: 2+1 quantum gravity,
  {\it Phys. Lett.} B272 (1991) 213-6
\item{[27]} J.E. Nelson and T. Regge: Invariants of 2+1 gravity,
 {\it Comm. Math. Phys.} 155 (1993) 561-8
\item{[28]} J.E. Nelson and T. Regge: Quantisation of 2+1
  gravity for genus 2, Torino preprint DFTT-54-93, gr-qc/9311029
\item{[29]} P. Peld\`an: Large diffeomorphisms in (2+1)-quantum
  gravity on the torus, preprint Penn State Univ. CGPG-95/1-1,
  gr-qc/9501020
\item{[30]} E. Martinec: Soluble systems in quantum gravity, {\it
  Phys. Rev.} D30 (1984) 1198-204
\item{[31]} A. Hosoya and K. Nakao: (2+1)-dimensional pure gravity
  for an arbitrary closed initial surface, {\it Class. Quant. Grav.}
  7 (1990) 163-76
\item{[32]} V. Moncrief: Reduction of the Einstein equations in 2+1
  dimensions to a Hamiltonian system over Teichm\"uller space, {\it
  J. Math. Phys.} 30 (1989) 2907-14
\item{[33]} Y. Fujiwara and J. Soda: Teichm\"uller motion of
  (2+1)-dimensional gravity with the cosmological constant, {\it
  Prog. Theor. Phys.} 83 (1990) 733-48
\item{[34]}  A. Hosoya and K. Nakao: (2+1)-dimensional quantum
  gravity, {\it Prog. Theor. Phys.} 84 (1990) 739-48
\item{[35]} R. Puzio: On the square root of the Laplace-Beltrami
  operator as a Hamiltonian, {\it Class. Quant. Grav.} 11 (1994)
  609-20
\item{[36]} S. Carlip: (2+1)-dimensional Chern-Simons gravity as a
  Dirac square root, {\it Phys. Rev.} D45 (1992) 3584-90, Err.:
  ibid., D47 (1993) 1729
\item{[37]} S. Carlip: Modular group, operator ordering, and time in
  (2+1)-dimensional gravity, {\it Phys. Rev.} D47 (1993) 4520-4
\item{[38]} S. Carlip: Observables, gauge invariance, and time in
  (2+1)-dimensional quantum gravity, {\it Phys. Rev.} D42 (1990)
  2647-54
\item{[39]} V. Moncrief: How solvable is (2+1)-dimensional Einstein
  gravity?, {\it J. Math. Phys.} 31 (1990) 2978-82
\item{[40]} A. Anderson: Unitary equivalence of the metric and
  holonomy formulations of the (2+1)-dimensional quantum gravity on
  the torus, {\it Phys. Rev.} D47 (1993) 4458-70
\item{[41]} S. Carlip: Notes on the (2+1)-dimensional Wheeler-DeWitt
  equation, {\it Class. Quant. Grav.} 11 (1994) 31-9
\item{[42]} M. Visser: Canonically quantized gravity: Disentangling
  the super-Hamiltonian and supermomentum constraint, {\it Phys.
  Rev.} D42 (1990) 1964-72
\item{[43]} V. Moncrief: Recent advances in ADM reduction, in: {\it
  Directions in General Relativity}, vol.1, ed. B.L. Hu et al., Cambridge
  University Press, 1993, 231-43

\end